\documentstyle[aps,prb,epsf,twocolumn,epsfig,floats]{revtex}

\begin{document}
\draft



\wideabs{

\title{The Frequency Dependent Conductivity of Electron Glasses}
\author{E. Helgren, N. P. Armitage and G. Gr\"{u}ner}
\address{Dept. of Physics and Astronomy, University of California Los Angeles, Los Angeles, CA 90095}

\date{\today}
\maketitle
\begin{abstract}

Results of DC and frequency dependent conductivity in the quantum
limit, i.e. $\hbar \omega > k_{B}T$, for a broad range of dopant
concentrations in nominally uncompensated, crystalline phosphorous
doped silicon and amorphous niobium-silicon alloys are reported.
These materials fall under the general category of disordered
insulating systems, which are referred to as electron glasses.
Using microwave resonant cavities and quasi-optical millimeter
wave spectroscopy we are able to study the frequency dependent
response on the insulating side of the metal-insulator transition.
We identify a quantum critical regime, a Fermi glass regime and a
Coulomb glass regime. Our phenomenological results lead to a phase
diagram description, or taxonomy, of the electrodynamic response
of electron glass systems.

\end{abstract}
} 

\section{Introduction}

Disordered insulating systems have garnered much interest from
theorists and experimentalists alike over many years. A plethora
of results exist on the low energy physics of these disordered
insulating systems from DC conductivity experiments, but it is
only recently that investigations of the $T \rightarrow 0$
frequency dependent response have been performed even though
theoretical predictions have existed for many years. We have
measured the DC conductivity as well as the complex frequency
dependent conductivity, the latter being measured in the
quantum-limit, i.e. $\hbar \omega > k_{B}T$ for a few different
disordered insulating systems. Measurements were performed on
materials that fall on the insulating side of the metal insulator
transition (MIT), with a major focus on the effects of
electron-electron interactions. Theoretical predictions concerning
the charge transport in systems with carriers localized in the
Anderson sense exist both for systems without electron-electron
interactions, these materials being referred to as Fermi glasses
\cite{Mott}, as well as for systems that include the effects of
the Coulomb interaction, these materials being referred to as
Coulomb glasses \cite{ES85}. The majority dopant concentration of
doped crystalline semiconductors and amorphous metal-semiconductor
alloys can easily be varied using standard semiconductor growth
techniques, and hence these materials provide a perfect playground
to study the low energy electrodynamics of localized charge
carriers.

We focus our investigation on a broad region of phase space from
close the the quantum critical point, i.e. the critical dopant
concentration at which the MIT occurs, $x_{c}$, to deep within the
insulating regime. Recent work on the AC conductivity of Si:B
\cite{MLee01} and Si:P \cite{Helgren2002} have shown a crossover
from the Fermi glass to the Coulomb glass-like behavior
qualitatively consistent with the theoretical expectations, but in
all cases, the crossover is much sharper than predicted by theory.
There exist two relevant energy scales in this problem. The first
is the Coulomb interaction energy, $U$, which Efros and Shklovskii
(ES) predicted determines the crossover energy between
characteristic Fermi glass and Coulomb glass-like behavior
\cite{ES85}. An important result of the interactions is a
depletion in the density of states of the charge carriers about
the Fermi level termed a Coulomb gap $\Delta$. In a recent result,
Lee et al.\cite{MLee01} claim from their measurements on Si:B that
the crossover energy scale is determined by the Coulomb gap width,
$\Delta$, as opposed to the theoretically predicted Coulomb
interaction energy, $U$. Arguments based on the concentration
dependence of these two parameters obtained from measurements over
a broad range of dopant concentrations in another canonical
electron glass system, Si:P, have recently been published by us
casting doubt on this interpretation \cite{Helgren2002}. This
highlights the usefulness and importance of investigating a wide
range of majority dopant concentrations to fully understand these
disordered insulating systems.

The primary goals of this work then are to present new data and to
draw on results of previous work on other electron glass systems
\cite{Helgren2002,Helgren2001,Henderson} to look at the insulating
side of the MIT using AC conductivity data and a broad range of
dopant concentrations in order to come up with a
phenomenologically supported phase diagram outlining the general
classification scheme for the electrodynamic response of electron
glasses. DC conductivity data are presented as well and a side by
side comparison of the AC and DC data provides for unique insight.
Furthermore the AC data allows for a detailed comparison of the
electron glass regime with theoretical predictions. We find that
some of the predicted features are observed, such as the frequency
dependent conductivity power laws. In fact it is these powers that
provide the taxonomic information in the phase diagram to be
presented herein. We also note discrepancies that arise in all the
reported measurements, such as the sharper than predicted
crossovers from Fermi glass to Coulomb glass behavior.

\section{Background}

\subsection{The MIT as a Quantum Phase Transition}

The metal insulator transition is an archetypical quantum phase
transition (QPT). The insulating portion of phase space associated
with Fermi glass and Coulomb glass behavior borders the quantum
critical regime. The quantum critical region of phase space is
that region where quantum mechanical effects due to being in
proximity to the zero temperature quantum critical point manifest
themselves even at finite temperatures or frequencies.
Specifically the electrodynamic response in a disordered
electronic system, such as the lightly doped insulating
semiconductors studied in this work, should smoothly transform
from Fermi glass or Coulomb glass-like behavior into the
appropriate quantum critical dynamic response as the dopant
concentration is tuned and the MIT is approached.

For the MIT, the dopant concentration which is directly related to
the charge carrier concentration, $x$, is typically the tuning
parameter that determines whether at zero temperature the material
is a metal, i.e. has a finite DC conductivity, or an insulator,
i.e. the DC conductivity is zero. There exists necessarily then a
critical concentration, $x_{c}$, at which the phase transition
occurs thereby defining a boundary between insulating and metallic
ground states. The disorder induced metal-insulator transition and
the dynamical behavior near the critical point has been
extensively theoretically studied over the past years both with
and without electron-electron interactions
\cite{Wegner,Abrahams1979,Shapiro1981,McMillan}. A thorough review
has been written by Belitz and Kirkpatrick \cite{Belitz}.

The condition for observing quantum critical behavior versus
electron glass-like behavior on the insulating side of the MIT is
determined by the relative size of certain characteristic length
scales. The characteristic thermal length scale is
\begin{eqnarray}
    \ell_{T} \propto (k_{B}T)^{-1/z}
    \label{eq:InelasticLengthTemp}
\end{eqnarray}
and the frequency dependent length scale is
\begin{eqnarray}
    \ell_{\omega} \propto (\hbar\omega)^{-1/z}
    \label{eq:InelasticLengthOmega}
\end{eqnarray}
where $z$ is the dynamical scaling exponent. These lengths are
termed "dephasing" lengths by Sondhi et al. \cite{Sondhi}. The
localization length, a measure of how far the electronic
wavefunction extends, is the other important length scale. It
diverges as the transition is approached and can be written in
terms of concentration as
\begin{eqnarray}
    \xi \propto |x - x_{c}|^{-\nu}
    \label{eq:localizationLength}
\end{eqnarray}
where $\nu$ is the localization length exponent.

\begin{figure}[tbh]
\centerline{\epsfig{figure=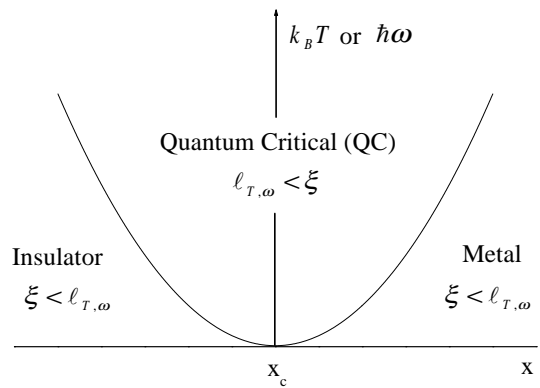,width=8cm}} \vspace{.2cm}
\caption{A schematic of the crossover diagram close to a quantum
critical point showing the regime boundaries between quantum
critical, metallic and insulating behavior. The horizontal axis
represents increasing concentration and the vertical axis
represents either the temperature or frequency energy scale. The
crossover from QC to either metallic or insulating behavior is
determined by when the length scales are equivalent, i.e.
$\ell_{T,\omega}\approx \xi$. \label{fig:mit}}
\end{figure}

At finite temperature or frequency an intermediate regime exists
between the metallic and insulating regimes that displays quantum
critical (QC) behavior as shown in Figure \ref{fig:mit}. In the QC
regime the localization length is greater than the dephasing
length scales, $\ell_{T,\omega}<\xi$. At the crossover from QC to
the insulating regime this can be thought of heuristically as the
electron scattering within a length scale that is shorter than the
width of the wavefunction envelope that defines how localized that
electron is. In other words the electron can undergo scattering
akin to normal transport before "realizing" that it is localized.
On the metallic side, the relevant length scale is commonly taken
to be the screening length. In the opposite limit,
$\xi<\ell_{T,\omega}$ the system exhibits metallic or insulating
behavior. For instance in the insulating regime, the material will
display the appropriate hopping conduction consistent with the DC
conductivity or the photon assisted frequency dependent
conductivity. A schematic representation of the phase diagram
showing the regime boundaries is shown in Figure \ref{fig:mit}.

From equations (\ref{eq:InelasticLengthTemp}),
(\ref{eq:InelasticLengthOmega}) and (\ref{eq:localizationLength})
we can determine a functional form, in terms of concentration, for
the crossover between the QC and the metallic or insulating
regimes by setting the dephasing length equal to the relevant
length scale, $\ell_{T,\omega}\approx \xi$. In so doing we find
\begin{eqnarray}
    \hbar \omega_{QC} \propto |x - x_{c}|^{z\nu}.
    \label{eq:CriticalSigmaac}
\end{eqnarray}
In this work the approach to criticality from the insulating side
of the MIT for two types of disordered insulators, Si:P and NbSi,
will be compared and discussed.\newpage

\subsection{DC Variable Range Hopping and \\the Coulomb Gap}

Mott first noted that at low temperatures in a lightly doped
disordered semiconductor a form of DC conduction unique from
nearest-neighbor hopping would occur \cite{Mott1968}. It is
commonly referred to as variable range hopping (VRH). Mott's
derivation did not include electron-electron interactions and thus
Mott-type hopping describes the variable range hopping mechanism
in a non-interacting Fermi glass \cite{Mott}. In Mott's theory:
\begin{equation}
    \sigma_{dc} \propto exp[-(T_{0}/T)^{1/(d+1)}]
    \label{eq:Mottdc}
\end{equation}
where d is the effective spatial dimension for the hopping
electrons. The characteristic Mott temperature is given by
\begin{equation}
    T_{0} = \beta / k_{B} N_{0} \xi ^{d}.
    \label{eq:T0_Mott}
\end{equation}
It is the characteristic energy scale of the level spacing in a
volume $\xi^{d}$, and dependent on the density of states (DOS),
$N_{0}$. Here $\beta$ is a constant of order one.

The above derivation was based on the assumption of an approximately constant density of states near the Fermi
level, but as pointed out by Pollak \cite{Pollak1970} and Ambegaokar \cite{Ambegaokar}, this assumption is invalid
when correlation effects are included. Upon including electron-electron interactions one finds that the Coulombic
repulsion is responsible for opening up a gap at the Fermi energy, and that the functional form for the DC
conductivity takes on an alternate temperature dependence. This gap, termed the Coulomb gap by Efros and Shklovskii
\cite{ES75} is given by
\begin{eqnarray}
    \Delta = \frac{e^{3} N_{0}^{1/2}}{\varepsilon_{1}^{3/2}},
\label{eq:GapWidth}
\end{eqnarray}
where $\varepsilon_{1}$ is the full dielectric constant. The
Coulomb gap is a "soft" gap in the density of states near the
Fermi level created by the Coulomb interaction between localized
electrons. (The term soft is used because the density of states
vanishes only at the Fermi energy.)

Efros and Shklovskii followed Mott's derivation for the DC
temperature dependent VRH conductivity, but instead of using a
constant density of states around the Fermi level, they
incorporated the Coulomb gap and therefore necessarily took
correlation effects into account. The appellation Coulomb glass
describes an electron glass system which incorporates Coulombic
interactions. In so doing ES found the following form for the DC
temperature dependent VRH conductivity
\begin{eqnarray}
    \sigma_{dc} \propto exp [-(T'_{0}/T)^{1/2}]
    \label{eq:ESdc}
\end{eqnarray}
where the characteristic ES temperature  is given by
\begin{eqnarray}
    T'_{0} \cong 2.8 e^{2} / k_{B} \varepsilon_{1} \xi.
    \label{eq:T0_ES}
\end{eqnarray}
The prefactor 2.8 is calculated from percolation threshold models
\cite{Nguyen,Efros1979b}. Note that the power in Eq.
(\ref{eq:ESdc}) is $\frac{1}{2}$ in all dimensions unlike in
Mott's formalism.

For the systems we are studying, any given sample with a
particular dopant concentration may possibly exhibit more than one
type of behavior. Depending on the energies at which one is
probing the system one might observe behavior consistent with
Fermi glass, Coulomb glass or QC-like behavior. Therefore, by
investigating a broad range of frequency or temperature space, our
experimental window might be broad enough to see crossovers from
one type of behavior to another.

The most indicative piece of experimental evidence that most
authors cite as to whether the materials being studied are Fermi
glasses exhibiting Mott VRH or Coulomb glasses exhibiting ES VRH
is the DC conductivity. The vicissitudes of using this
experimental data to claim a material is a Coulomb or Fermi glass
lie in the fact that only below a certain temperature does a
material enter the VRH regime, whereas above this cutoff
temperature other forms of activated hopping occur. Thus only a
small portion of low temperature data will show any semblance of
VRH behavior, and in three dimensions, a small section of
temperature dependent data may fit a power law of $\frac{1}{2}$ or
$\frac{1}{4}$ almost equally as well making it difficult to say
with any certainty which mechanism of charge excitation is
occurring. Measurements of the $T \rightarrow 0\;K$ frequency
dependent conductivity then allows for new insight into the MIT
and the electrodynamics of electron glasses. Measurements of the
DC conductivity on the same samples as those on which the AC
conductivity measurements were taken offers a consistency check
with previous investigations and allows for comparison of salient
features.

\subsection{AC Photon-assisted Hopping}

Mott proposed a form for the $T = 0 \; K$ photon assisted
frequency dependent hopping conductivity for the case of
non-interacting electrons based on the theory of resonant
absorption using a simple one electron model of a disordered
system, commonly referred to as the pair approximation
\cite{Mott,TanakaFan}. The resulting real part of the AC
conductivity, $\sigma_{1}$, in Mott's formalism is as follows:
\begin{equation}
    \sigma_{1} = \alpha
    e^{2}N_{0}^{2}\xi^{5}\hbar\omega^{2}[ln(2I_{0}/\hbar\omega)]^{4}.
    \label{eq:Mottac}
\end{equation}
Here $\alpha$ is some constant close to one, and $I_{0}$ is the
pre-factor of the overlap integral (commonly taken to be the Bohr
energy of the dopant \cite{Shklovskii}). Most notably we see that
in Mott's formalism, the conductivity has an approximately
quadratic frequency dependence, plus logarithmic corrections.

\begin{figure}[tbh]
\centerline{\epsfig{figure=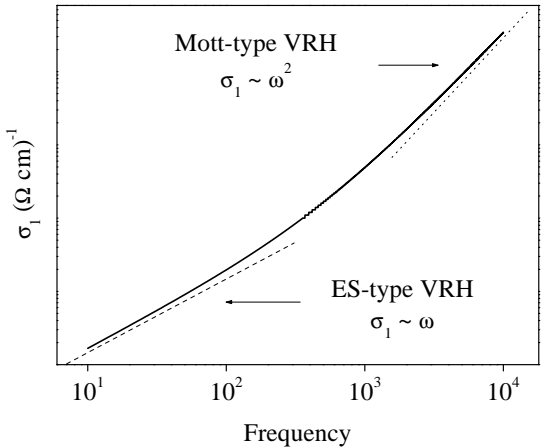,width=8cm}}
\vspace{.2cm} \caption{A schematic of the crossover function from
Mott to ES-type AC conduction set forth by Efros and Shklovskii
and defined in Eq. (\ref{eq:ESxover}) displayed on logarithmic
axes. The vertical axis is conductivity with arbitrary units and
the horizontal axis has been set to show the crossover occurring
in the millimeter-wave range if the units are taken to be GHz. The
dashed and dotted straight lines show the asymptotically
approached linear and quadratic dependencies of the crossover
function. \label{fig:ModelCrossover}}
\end{figure}

By taking into account the mean Coulomb interaction between two
sites forming a resonant pair
\begin{eqnarray}
    U(r_{\omega}) = e^{2}/\varepsilon_{1} r_{\omega}
    \label{eq:U}
\end{eqnarray}
where $r_{\omega} = \xi[ln(2I_{0}/\hbar\omega)]$ is the most
probable hop distance between pairs, Efros and Shklovskii derived
the real part of the AC conductivity to be:
\begin{equation}
    \sigma_{1} = \alpha e^{2}N_{0}^{2}\xi^{5}\omega[ln(2I_{0}/\hbar\omega)]^{4}[\hbar\omega +
    U(r_{w})].
    \label{eq:ESxover}
\end{equation}
This formula takes on a different frequency dependence in two
limiting cases. In the case where the photon energy is greater
than the Coulomb interaction energy, $\hbar\omega >
U(r_{\omega})$, one retrieves the same approximately quadratic
frequency dependence in Mott's formula (\ref{eq:Mottac}), and here
the Coulomb glass is indistinguishable from the Fermi glass in so
far as the frequency dependent conductivity is concerned. For the
alternate case i.e. when the photon energy is smaller than the
Coulomb interaction energy, $\hbar\omega < U(r_{\omega})$, the
conductivity of a Coulomb glass will show an approximately linear
dependence on frequency, plus logarithmic corrections. Model data
for this functional form is shown in Figure
\ref{fig:ModelCrossover} showing that the crossover is predicted
to occur over many decades of frequency.

As for the DC case, one further ramification of including electron-electron interactions is the depletion of the
single particle density of states around the Fermi level, i.e. the Coulomb Gap $\Delta$. In the special case when
$\Delta > U(r_{\omega}) > \hbar\omega$ Efros and Shklovskii calculated the real part of the frequency dependent
conductivity and obtained
\begin{equation}
    \sigma_{1} \approx \frac{1}{10}\frac{\omega \varepsilon_{1} }{ln(2I_{0}/\hbar\omega)},
    \label{eq:ESinGap}
\end{equation}
so that the dependence is slightly super-linear, assuming a frequency-independent dielectric constant, and
independent of the localization length, which would indicate a very weak concentration dependence.

Thus, by using the frequency dependent conductivity one has a
unique means to determine the relevant energy scales between
interacting and non-interacting behavior. Several recent
experiments \cite{MLee01,HellmanDynes2000} have shown a possible
crossover from interacting to non-interacting behavior in the same
material. Some of this experimental data has been interpreted to
indicate that the Coulomb gap width energy scale, certainly
another germane energy scale in the system, sets the crossover
energy scale in opposition to the theoretical prediction that the
Coulomb interaction energy dictates the crossover from Fermi glass
to Coulomb glass-like behavior. These arguments though were based
on measurements covering a limited range of dopant concentrations.
Therefore the frequency dependent conductivity measured in the low
energy excitation range for a broad range of concentrations can
provide useful information about the crossover energy scale, be it
the Coulomb gap width or the Coulomb interaction energy.

\section{Experiments}

Experiments were performed on nominally uncompensated phosphorous
doped silicon samples that were cut from a boule grown by Recticon
Enterprises Inc. to a specification of 5 cm in diameter with a
majority dopant gradient along the axis. This boule, grown using
the Czochralski method, was subsequently sliced into 1 mm thick
wafers. By the nature of the Czochralski method for growing a
crystal, there is an inherent gradient in the majority dopant
concentration along the axis of the boule. A greater range of
variation in this dopant concentration along the axis of the boule
can be achieved by continuously adding progressively more dopant
of choice, in our case phosphorous, to the molten silicon as the
ingot is being drawn from the liquid mixture. As summarized in the
first column in Table \ref{tab:SiP UCLA}, we obtained samples
spanning a majority dopant concentration from $1.37\times 10^{18}
\; cm^{-3}$ to $2.98 \times 10^{18} \; cm^{-3}$. The middle column
lists the ratios of the samples as a percentage of the dopant
concentration in each sample relative to the critical
concentration at which the MIT occurs, $x_{c} = 3.5 \times 10^{18}
\; cm^{-3}$. The final column lists the room temperature
resistivity of each sample in $(\Omega \; cm)$ measured using an
ADE 6035 resistivity gauge, which can be directly related to the
dopant concentration, knowing the ilk of majority dopant in the
silicon via a phenomenological scale set forth by Thurber et al.
\cite{Thurber}.

\begin{table}[hbt]
\centering\begin{tabular}{c|c|c}

    $x$ $\times 10^{18}\; cm^{-3}$ & Percent of Critical: $\frac{x}{x_{c}}$
    & $\rho_{dc}(300\;K) \;(\Omega \;cm)$ \\
    &  &   \\
    \hline \hline
    2.98 & 85\% & 0.0133  \\\hline
    2.66 & 76\% & 0.0142  \\\hline
    2.42 & 69\% & 0.0150  \\\hline
    2.17 & 62\% & 0.0159  \\\hline
    1.96 & 56\% & 0.0168  \\\hline
    1.74 & 50\% & 0.0179  \\\hline
    1.56 & 45\% & 0.0189  \\\hline
    1.37 & 39\% & 0.0202

\end{tabular}
\caption{The material parameters for eight Si:P samples obtained from a boule grown by Recticon Enterprises Inc.
The left hand column lists the concentration of the majority dopant phosphorous in each sample. The middle column
lists the ratio as a percentage of the dopant concentration in each sample relative to the critical concentration
of the MIT, namely $x_{c} = 3.5 \times 10^{18} \; cm^{-3}$. The right hand column lists the value of the room
temperature DC resistivity value for each sample.} \label{tab:SiP UCLA}
\end{table}

Previously reported results from another electron glass system,
namely amorphous NbSi, will be drawn upon when discussing the
general features of the electron glass phase diagram. These
samples were created by sputtering onto sapphire substrates, and
both millimeter-wave and DC conductivity measurements were
performed, analogous to those done on Si:P. More detailed sample
specifications for these amorphous metal-semiconductor alloys and
experimental results can be found in the literature
\cite{Helgren2001}.

It is important to perform DC conductivity measurements on the
same Si:P samples for which the AC measurements were done for a
number of reasons. First, the DC measurements offer a consistency
check to previous experimental results. Also, one can compare
important parameters, such as the localization length, as
determined separately from the DC and AC data. These measurements
also provided a low temperature value of the DC conductivity to
which one can normalize the real part of the microwave resonant
cavity complex conductivity. For example, at approximately $25$ K
the real part of the conductivity measured at the center
frequencies of the resonant cavities, namely $35 \; GHz$ and $60
\; GHz$, should be dominated by the thermally driven charge
transport mechanism that determines the DC conductivity.

\begin{figure}[tbh]
\centerline{\epsfig{figure=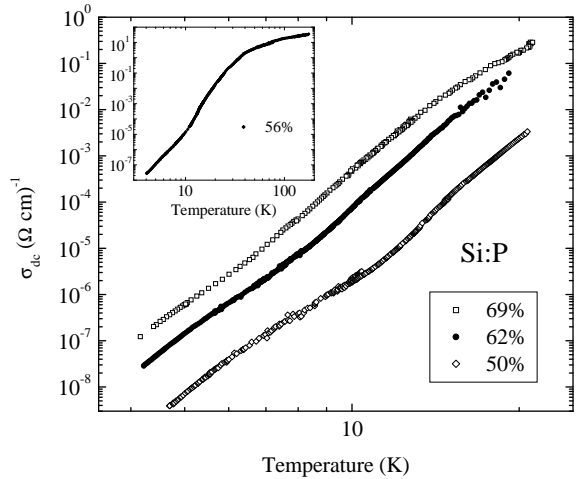,width=9cm}}
\vspace{.2cm} \caption{The DC conductivity versus temperature on
logarithmic scales for a series of Si:P samples. At high
temperatures one can notice a crossover to activated
Boltzmann-type behavior. Also noticeable is a crossover
temperature, $T_{x}$, in the middle of the displayed temperature
range that decreases with increasing dopant concentration.
\label{fig:SiP all etched DC}}
\end{figure}

DC conductivity measurements were performed on the Si:P samples using a standard four probe measurement technique
from liquid $^{4} He$ temperature up to approximately $100$ K. Figure \ref{fig:SiP all etched DC} shows data from
three samples from $4.2$ K up to $20$ K. At the higher temperatures shown in the main portion of the figure for the
the 69 \% sample, and also evident in the inset plot of the 56 \% sample measured up to approximately 200 K, the
curves show a rounding off indicative of a conduction process consistent with a Boltzmann type activation. Also
evident in the middle of the displayed temperature range is a kink in each of the sample's data indicative of a
crossover from one form of activation to another. Visually the crossover temperature, $T_{x}$, for each of the
samples decreases with increasing dopant concentration. This would be consistent with a crossover from Mott to
ES-type VRH, but other experimental evidence, to be discussed presently, indicates that Mott and ES-type VRH
theories do not correctly describe the upper and lower pieces of the DC conductivity data shown in Figure
\ref{fig:SiP all etched DC}.

\begin{figure}[tbh]
\centerline{\epsfig{figure=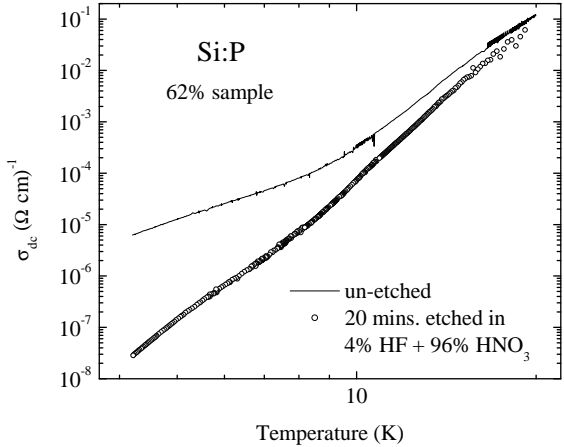,width=9cm}}
\vspace{.2cm} \caption{The DC conductivity versus temperature on a
logarithmic scale for the 62\% Si:P sample both prior to and after
etching the sample. \label{fig:SiP etched vs unetched}}
\end{figure}

Figure \ref{fig:SiP etched vs unetched} shows the measured DC
conductivity from the 62\% sample both prior to and after etching
in a $4\% \; HF + 96\% \; HNO_{3}$ solution for 20 minutes. Unless
otherwise specified all AC and DC conductivity measurements
reported herein were performed on etched Si:P samples. As first
pointed out by Ootuka et al. \cite{Kobayashi1980b}, a crystalline
semiconductor sample sliced from a boule using a diamond saw will
create surface defects. As indicated in the figure, at low
temperatures the surface defects seem to provide a channel for
conduction with less resistance than the surface defect-free bulk
sample, increasing the overall measured bulk conductivity. DC
conductivity measurements on samples etched for 20 minutes and 2
hours showed no difference indicating that 20 minutes of etching
sufficed to remedy the effects of surface defects, consistent with
the findings of Ootuka et al. \cite{Kobayashi1980b}. Regardless of
the presence of surface defects altering the relative magnitude of
the measured bulk conductivity, the kink in the data indicating a
crossover temperature, $T_{x}$, is still clearly present in both
sets of data.

Both components of the complex frequency dependent conductivity
were determined from transmission measurements using a technique
called quasi-optical spectroscopy. Quasi-optical spectroscopy has
been utilized and partially developed by our research group to
investigate the electrodynamic response of materials in the
millimeter to sub-millimeter wave range (i.e. the terrahertz
frequency regime) \cite{Schwartz,Gruner}. Historically the
millimeter and sub-millimeter wave spectral range, higher in
frequency than standard radio-frequency (RF) techniques and lower
in frequency than common infrared (IR) techniques, has not been
studied, leaving a gap in the knowledge of the optical response of
physically intriguing systems. This spectral range corresponds
with the range of energies wherein electron-electron interactions
are predicted to occur in disordered systems. Also experiments in
this spectral range can easily be done in the quantum limit,
$\hbar\omega > k_{B}T$, with standard cryogenic techniques, i.e a
pumped liquid $^{4} He$ optical cryostat (recall that $1\;K \simeq
20 \; GHz$).

The quasi-optical transmission measurement technique takes
advantage of broadband sources of coherent electromagnetic
radiation known as Backward Wave Oscillators (BWO). The frequency
of the output radiation from these sources can be swept by
adjusting the applied potential. The resulting output power
spectrum is extremely reproducible Recording the power received at
the detector for 1) the sample and then 2) a blank or free space,
and taking their ratio provides the transmission data as a
function of frequency. The striking feature of the transmission
spectra is the periodically spaced peaks known as Fabry-Perot
resonances which occur whenever the thickness of the sample is
equal to an integer number of half wavelengths inside of the
sample. The transmission can be recorded and the resonant peak's
center frequencies and heights can be determined. From these
measured parameters one can directly determine the real and
imaginary components of the complex conductivity.

\begin{figure}[tbh]
\centerline{\epsfig{figure=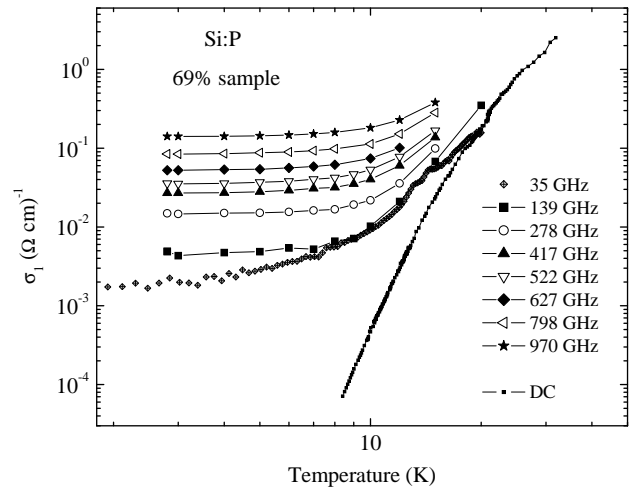,width=9cm}}\caption{Both
the DC conductivity and the real component of the complex
conductivity at various constant frequencies versus temperature on
logarithmic scales for the 69\% sample.\label{fig:SiP DC and AC}}
\end{figure}

Below a certain temperature, it was found that the Fabry-Perot
resonance transmission traces did not change. This necessarily
means that below some temperature the real and imaginary
components of the complex conductivity, become temperature
independent. At low temperatures thermal contractions, which can
affect both the peak heights and center frequencies become
negligible as well. Figure \ref{fig:SiP DC and AC} shows this
trend by plotting the conductivity of the 69\% sample versus
temperature for various constant frequencies with the DC
conductivity. It is clear that at high temperatures the AC
conductivity curves approach and follow the temperature dependent
DC conductivity, and the greater the frequency of a particular
curve, the greater the temperature at which this occurs. It is
also clear that the iso-frequency curves do indeed flatten out at
low temperatures, indicating that the $T=0\;K$ photon-assisted
hopping behavior is being observed. All of the other samples
showed a similar flattening out of the AC conductivity at low
temperatures.

\begin{figure}[tbh]
\centerline{\epsfig{figure=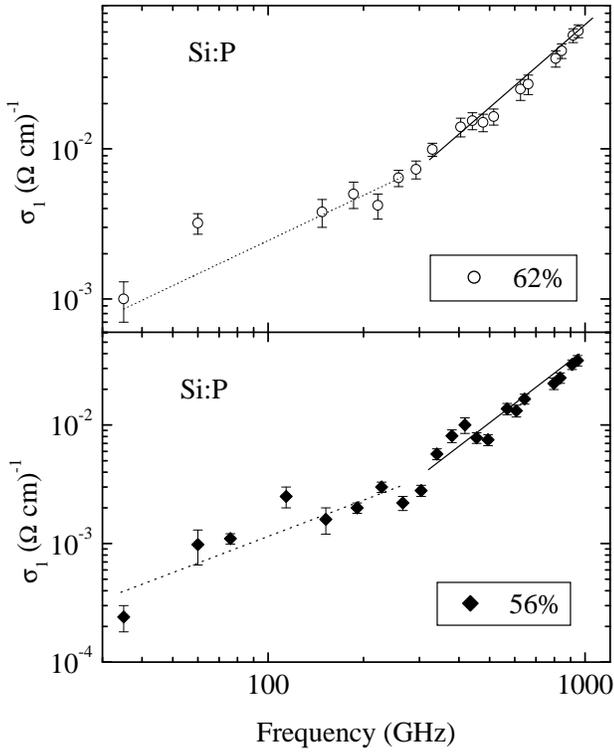,width=9cm}}
\caption{The measured 2.8 K value of the real part of the complex
frequency dependent conductivity for the Si:P 62\% and 56\%
samples plotted versus frequency on logarithmic axes displaying a
crossover in the type of conduction mechanism. The dashed line is
a fit to the lower portion of the data and follows a nearly linear
power law. The solid line through the upper portion of the data is
a fit of that data and follows a nearly quadratic power law.
\label{fig:SiP xover 62 and 56}}
\end{figure}

Using DC data alone, evidence of the effects of electron-electron interactions, namely a crossover from Mott to ES
type VRH is questionable at best. The range of temperature across which the various conduction processes might be
occurring are minimal and it is difficult to have confidence that a fit to Mott-type $T^{-1/4}$ behavior or ES-type
$T^{-1/2}$ behavior is correct. The distinguishing features of an interacting versus a non-interacting disordered
system are much easier to tell apart when investigating the frequency dependent conductivity though. Specifically,
linear and quadratic power law are clearly discernable and evidence for such a crossover in the frequency dependent
conductivity spectrum is clearly seen as shown in Figures \ref{fig:SiP xover 62 and 56} and \ref{fig:SiP xover 45
and 39}.

\begin{figure}[tbh]
\centerline{\epsfig{figure=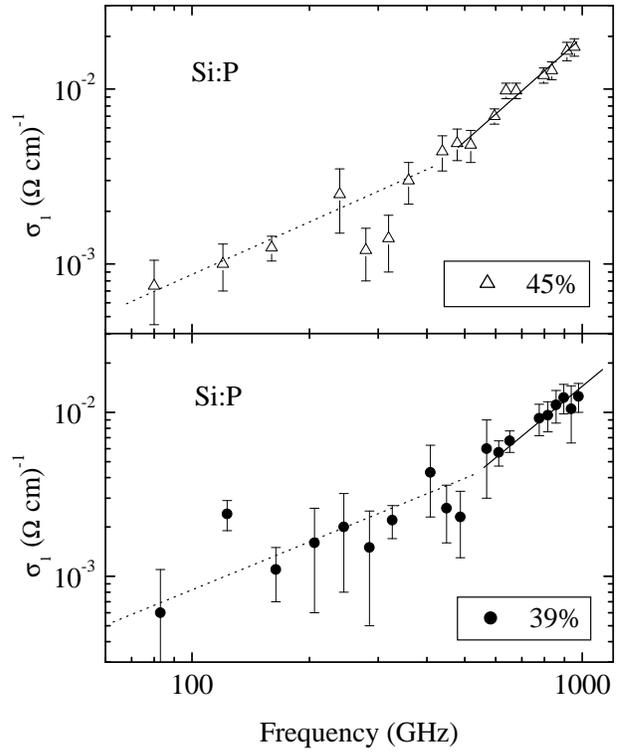,width=9cm}}
\caption{The measured 2.8 K value of the real part of the complex
frequency dependent conductivity for the Si:P 45\% and 39\%
samples plotted versus frequency on logarithmic axes displaying a
crossover in the type of conduction mechanism. The dashed line is
a fit to the lower portion of the data and follows a nearly linear
power law. The solid line through the upper portion of the data is
a fit of that data and follows a nearly quadratic power law.
\label{fig:SiP xover 45 and 39}}
\end{figure}

\section{Analysis and Results}

\subsection{DC Conductivity}

\begin{figure}[tbh]
\centerline{\epsfig{figure=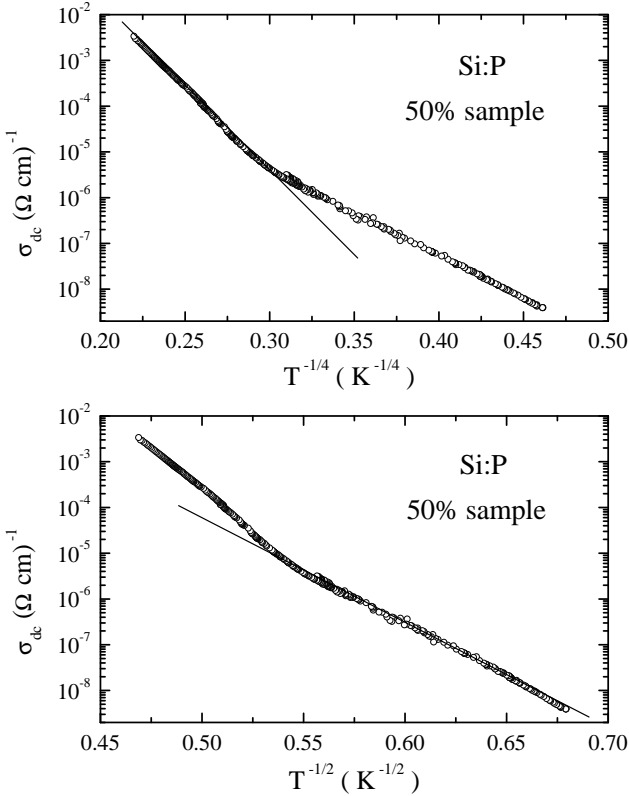,width=9cm}}\caption{The
DC conductivity for the 50\% Si:P sample plotted versus $T^{-1/4}$
in the upper panel and plotted versus $T^{-1/2}$ in the lower
panel. \label{fig:SiP dc fits}}
\end{figure}

Much of the story that the DC conductivity can tell is summarized
in Figure \ref{fig:SiP dc fits} which shows the DC conductivity of
the 50\% Si:P sample. The data in the upper panel is plotted on a
$T^{-1/4}$ axis in order to assist visual recognition of Mott type
VRH behavior. The lower panel shows the sample data plotted on a
$T^{-1/2}$ axis in an attempt to elucidate any region of data that
might be following ES type VRH behavior. When plotted on any
inverse temperature scale, the left hand side represents the high
temperature region, and it is this higher temperature region, up
to approximately 25 K, that has been fit to a linear function in
the upper panel. However, closer inspection of Figure \ref{fig:SiP
dc fits} reveals that  in the upper panel, although the higher
temperature data has been fitted to $T^{-1/4}$ behavior, the lower
portion seems to be linear as well indicating a possible
$T^{-1/4}$ dependence as well on this semi-log plot. In the lower
panel of Figure \ref{fig:SiP dc fits}, the lower temperature data
is fitted to a linear function, possibly indicating that in this
region, i.e. below $T_{x} \approx 12\:K$, the data might well be
described as following ES type behavior. As in the upper panel,
the other region of data, in this case the higher temperature
data, shows a somewhat linear dependence at least over a finite
range of temperature.

That the upper and lower portions of the data seem linear on
either scale summarizes the vicissitudes of trying to determine
Mott or ES type behavior from the DC data alone. From Figure
\ref{fig:SiP dc fits} one sees that the Si:P 50\% sample has a
crossover temperature of approximately 12 K, whereupon the DC
conductivity begins to follow a different dependence. It is
unclear if this temperature sets the energy scale at which a
crossover from Mott to ES type VRH occurs, but nevertheless it is
clearly an important energy scale in this model disordered system.

Further analysis shows that fitting the high temperature and low
temperature portions of $\sigma_{dc}$ respectively to Mott and
ES-type functional forms, across our experimentally accessible
temperature range, is incorrect. The characteristic Mott
temperatures, $T_{0}$, determined for our Si:P samples is of the
order $10^{7} \; K$, a ridiculously large value. The ES
characteristic temperatures, $T'_{0}$, determined by fitting the
lower temperature data are more reasonable but still extremely
large, of the order $10^{3} \; K$. Compared to the characteristic
Mott and ES temperatures measured by Dai et al.
\cite{Sarachik1991} and Hornung et al. \cite{Lohneysen2000}, our
values seem extremely high and unphysical, even though their
materials were of a different composition, or fell in a different
range of dopant concentrations.

Further evidence that the Mott and ES functional forms are
inappropriate can be gleaned from analysis of the localization
lengths from the characteristic temperatures. Using the
relationship for the ES characteristic temperature, i.e. Eq.
\ref{eq:T0_ES} and the measured values of $T'_{0}$, the
localization lengths for the various Si:P samples within our
doping range come out to be less than $2 \; nm$, not an
unreasonable length scale as the interatomic spacing in the
diamond structure lattice of Si is 2.35 \AA, but an order of
magnitude smaller than the $20 \; nm$ found from analyzing the AC
conductivity data, to be presented in the following section. With
an appropriate estimate of the noninteracting density of states
$N_{0} (E_{F})$, an estimate of the localization length can also
be determined from $T_{0}$. For example the 62\% sample, with a
$T_{0} = 39\times10^{6}\;K$, estimating $N_{0} (E_{F})$ to be the
ratio of the dopant atomic density $x = 2.173\times10^24\;m^{-3}$
to an appropriate bandwidth, $0.1\;eV$, gives a localization
length of approximately 6 \AA, a factor of 4 smaller than the one
determined from $T'_{0}$. That the localization lengths do not
agree gives further evidence that 1) either our assumptions of
what regions of the DC conductivity correspond to Mott or ES type
VRH are incorrect or that 2) deep in the insulating regime
behavior different than that predicted for canonical Coulomb and
Fermi glasses exists.

The former case may well be true, (although evidence from the AC
conductivity measurements seem to indicate otherwise), and these
issues will be discussed presently, but first the latter issue
shall be addressed. In fact such a prediction that deep in the
insulating regime an alternate method of charge transfer should
exist is exactly what Hornung et al. \cite{Lohneysen2000} claim to
have observed in their DC conductivity measurements of Si:P. Among
other findings, these authors claim that below a majority dopant
concentration of $2.78\times10^{18}\;cm^{-3}$, (recall that $x_{c}
= 3.52 \times 10^{18}\;cm^{-3}$) the conductivity was better
described by an activation process with an activation energy
$E_{2}$ which was interpreted as a "hard" Hubbard gap.

Inspection of the $1.80\times10^{18}\;cm^{-3}$ and
$2.40\times10^{18}\;cm^{-3}$ dopant concentration data presented
in literature \cite{Lohneysen2000}, which were analyzed as an
activation process, show that the data spanned a very small
temperature range. Our 69\% sample corresponds to a dopant
concentration very similar to the $2.40\times10^{18}\;cm^{-3}$
dopant concentration sample in Ref. 29. Our DC conductivity
results extended to 4.2 K. As indicated in Figure \ref{fig:SiP all
etched DC} a kink in the curve is clearly evident, which occurs at
or somewhat below the lowest temperature indicated in the work
from Hornung et al. \cite{Lohneysen2000}. The change in functional
form that we observe is inconsistent with the presence of a hard
gap which would result in a single activated process as the
conduction mechanism. As we will show below, further evidence for
the absence of a hard gap comes from the fact that we observe
appreciable AC conductivity at frequency scales well below that of
the purported gap energy.

\subsection{AC Transport and Polarizability: The Crossover}

Some of the most stunning and germane results from these
investigations of electron-electron interactions in disordered
systems are best summarized visually in Figure \ref{fig:SiP xover
fitted} in which the measured real part of the complex
conductivity for the 69\% and 50\% samples are displayed as a
function of frequency. The solid lines are linear and quadratic
fits to the lower frequency and higher frequency data
respectively, qualitatively displaying the expected behavior for a
crossover from a Coulomb glass to a Fermi glass. The overlayed
dotted lines are fits to the data using Eq. (\ref{eq:ESxover}) and
clearly this formula provides only a rough guide. The fit was made
by forcing the linear portion to pass through the origin as well
as the low frequency data and leaving the prefactor of the
quadratic portion as a free parameter. The crossover between
linear and quadratic behavior is much more abrupt than the ES
function predicts and is observed over our entire doping range.
Such behavior was previously observed in an an analogous
disordered semiconductor system, Si:B, for two very closely spaced
doping levels near the MIT \cite{MLee01}.

\begin{figure}[tbh]
\centerline{\epsfig{figure=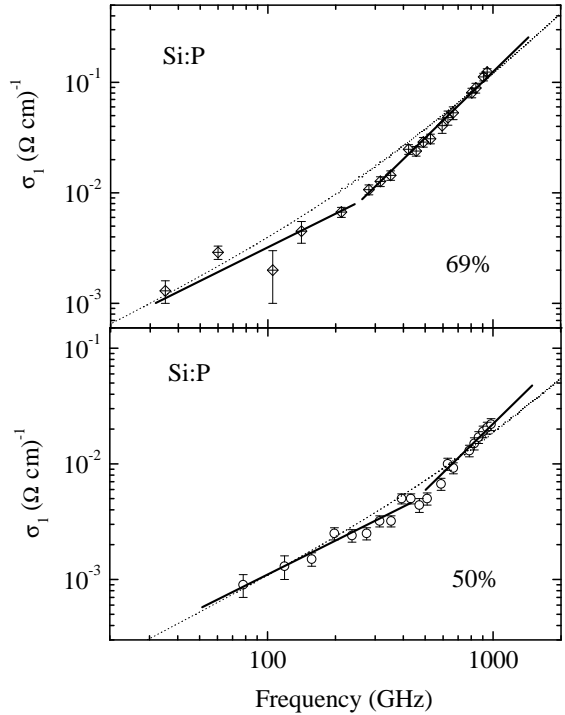,width=8cm}}\caption{The
measured real part of the complex conductivity for the 69\% and
50\% Si:P samples as a function of frequency. The solid lines are
linear and quadratic fits to the lower frequency and higher
frequency data respectively qualitatively displaying the expected
behavior for a crossover from a Coulomb glass to a Fermi glass.
The overlayed dotted lines are fits to the data using Eq.
(\ref{eq:ESxover}). \label{fig:SiP xover fitted}}
\end{figure}

\begin{figure}[tbh]
\centerline{\epsfig{figure=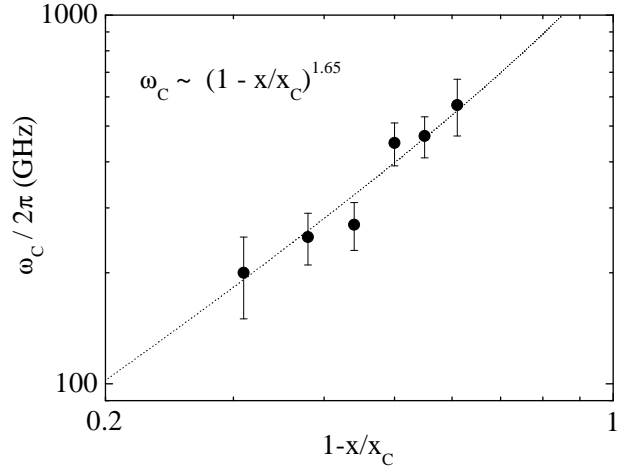,width=9cm}}\caption{The
experimentally determined crossover frequencies for the various
Si:P samples. The data is plotted against concentration in a
functional form $\left(1-\frac{x}{x_{c}}\right)^{\beta}$. We find
$\beta = 1.65$. \label{fig:SiP wc}}
\end{figure}

The sharpness of the crossover is interesting in its own right and
deserves further investigation as we do not here propose an
explanation. This crossover in the real part of the complex
conductivity was observed across our entire dopant concentration
range except for samples very close the MIT, where the crossover
presumably occurs below our experimental frequency range. The
experimentally determined crossover frequency, $\omega_{c}$,
plotted versus concentration in a functional form
$\left(1-\frac{x}{x_{c}}\right)$ is summarized in Figure
\ref{fig:SiP wc}.

The functional form in Figure \ref{fig:SiP wc}, namely a power law
in $1-\frac{x}{x_{c}}$, is chosen because both the localization
length, $\xi$, and the dielectric susceptibility of a sample near
the MIT, namely $4 \pi \chi = \varepsilon_{1} - \varepsilon_{Si}$,
are either predicted to or have been experimentally shown to
follow such a form. Fitting the crossover frequency to such a
functional form we find that $\omega_{c}$ is proportional to
$\left( 1 - \frac{x}{x_{c}}\right) ^{\beta}$ where $\beta = 1.65$.
By comparing the concentration dependencies of the relevant energy
scales to this experimentally determined power we will show that
the Coulomb interaction energy determines the crossover.

Recall that two energy scales have been set forth as being the
energy scale at which the crossover from Fermi glass to Coulomb
glass-like behavior is to occur. The original theoretical
framework set forth by Efros and Shklovskii \cite{ES85} has the
Coulomb interaction energy $U$ as the important energy scale. The
second relevant energy scale in these systems is the Coulomb gap
width, $\Delta$. The evidence that the Coulomb gap is the
pertinent energy scale of the system stems from an experiment
performed by Lee et al. \cite{MLee01}. This report claims that it
is not a gap in the single particle density of states, but rather
a "dressed" Coulomb gap in the multi-particle density of states
that determines the critical energy scale. Their claim, for which
there does exist a theoretical framework \cite{Pollakbook85}, is
that hopping phenomena occur due to multi-particle processes or
polaronic effects, but that a Coulomb gap still forms in the
density of states of these charge carrying entities. Nevertheless
it is not unreasonable to believe that the dressed Coulomb gap
should take a similar form to that of the ES single particle gap,
Equation (\ref{eq:GapWidth}), as both are presumably caused by the
same long range Coulomb interaction.

\begin{figure}[tb]
\centerline{\epsfig{figure=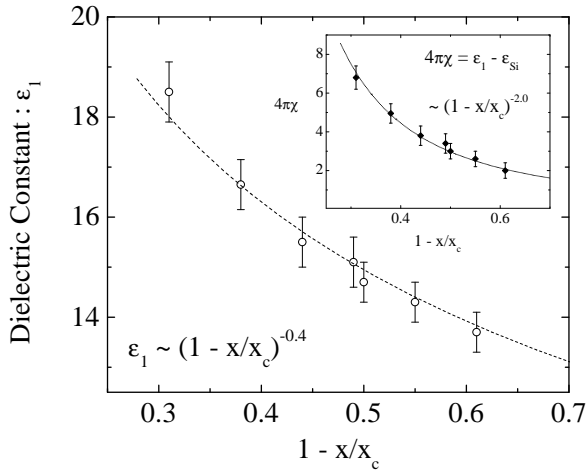,width=9cm}}\caption{The
$T=2.8\;K$ full measured dielectric constant $\varepsilon_{1}$ for
the various Si:P samples are plotted versus $1-\frac{x}{x_{c}}$.
The inset is that portion of the dielectric constant due to the
disordered interacting system itself, $4 \pi \chi$ obtained by
subtracting off the background dielectric constant
$\varepsilon_{Si}$ (of the host material silicon) from the full
dielectric constant $\varepsilon_{1}$. A clear divergence as a
function of concentration is seen. These measurements are
consistent with previous reports, and facilitate the analysis of
the crossover energy scale. \label{fig:SiP_Diel_Konst}}
\end{figure}

The dielectric constant enters into many of the germane formulae
describing the electrodynamics of these disordered insulating
systems. The effects of screening and polarization become more and
more relevant as the MIT is approached. Therefore it is of
fundamental importance to develop a theory that self-consistently
takes into account the effects of the imaginary component of the
frequency dependent conductivity, i.e. the dielectric constant. As
mentioned before, the immense advantage of our experimental
techniques is the fact that we were able to separately measure
both components of the complex conductivity. The full measured
dielectric constant, as determined at $T=2.8\;K$, which is
independent of frequency across our accessible frequency range (50
- 1000 $GHz$), is displayed in Figure \ref{fig:SiP_Diel_Konst}.
The inset shows that portion of the dielectric constant due to the
disordered interacting system itself, which we refer to as the
dielectric susceptibility of the disordered system, $4\pi\chi$,
obtained by subtracting off the background dielectric constant
$\varepsilon_{Si}$ (of the host material silicon) from the full
dielectric constant $\varepsilon_{1}$.

Measurements of the concentration dependent divergence of the
dielectric constant are not new but having independently
determined the values for our samples at these frequency ranges
allows for more accurate analysis of the crossover energy scales.
As a measure of confidence, when compared to previous experimental
data on the dielectric susceptibility portion, namely $4 \pi
\chi$, of the full dielectric constant for Si:P \cite{Hess}, which
was found to fit a form $\left( \frac{x_{c}}{x} - 1
\right)^{\zeta}$, with $\zeta =1.15$, we obtain from our data,
when plotted and fitted in a similar fashion, a power also very
close to unity, namely 0.96. However, measurements of the full
dielectric constant, $\varepsilon_{1}$, reported in the literature
on Si:B \cite{MLeePRB1999}, a disordered insulator analogous to
Si:P, were fitted to a functional form of $\left( 1 -
\frac{x}{x_{c}} \right)^{\zeta '}$, and the power was found to be
$\zeta ' = -0.71$. We obtain from our data, when plotted and
fitted in a similar fashion, a power for $\varepsilon_{1} \propto
\left( 1 - \frac{x}{x_{c}} \right)^{\zeta '}$ of $\zeta ' =
-0.38$, which is almost a factor of 2 different. This is perhaps
accounted for by the fact that the materials have different dopant
material. One must be cautious when analyzing the full dielectric
constant in this latter manner as is found sometimes in the
literature because there is a constant additive term due to the
background dielectric constant of the host material Silicon
$\varepsilon_{Si} = 11.7$ contained in the full dielectric
constant $\varepsilon_{1} = \varepsilon_{Si} + 4 \pi \chi$, and
only at dopant concentrations very close to critical does one
obtain $\varepsilon_{1} \approx 4 \pi \chi \gg \varepsilon_{Si}$.
Our subsequent analysis of the crossover energy between the
non-interacting Fermi glass behavior and the electron-electron
interaction dependent behavior of the Coulomb glass state is
carried out using the latter normalized concentration functional
form, $\left( 1 - \frac{x}{x_{c}} \right)$. It is therefore
important to note here that as displayed in the inset of Figure
\ref{fig:SiP_Diel_Konst}, the dielectric susceptibility analyzed
in this manner is found to obey the power law dependence, $ 4 \pi
\chi \propto \left( 1 - \frac{x}{x_{c}} \right)^{-2}$, which
presumably becomes the dominant power law dependence arbitrarily
close to the MIT as the host material's $ \varepsilon _{Si}$
becomes negligibly small. In fact such a trend can be seen in
$\varepsilon _{1}$ in Figure \ref{fig:SiP_Diel_Konst}, where at
higher concentrations, namely the left-hand side of the figure,
the slope tends to be steeper, as compared to the lower dopant
concentrations on the right-hand side of the plot.

Utilizing the measured values of the full dielectric constant for each sample one can now analyze the crossover
energy. We mention briefly that in our analysis we parameterize $\omega_{c}$ in terms of concentration. We
acknowledge that the crossover energy should not scale as a power law over the whole doping range, but in our
limited range modelling shows this to be an acceptable parameterization. Now let us first examine the Coulomb gap
width $\Delta$. The experimental data shows that the crossover frequency $\omega_{c}$ is proportional to $\left(1 -
\frac{x}{x_{c}} \right) ^ {\beta}$ where from our analysis $\beta = 1.65$, and the full dielectric constant
$\varepsilon_{1} \propto \left(1 - \frac{x}{x_{c}} \right) ^ {-0.4}$. Using these values and Eq. \ref{eq:GapWidth}
one finds that the DOS goes as $N_{0} \propto \left( 1 - \frac{x}{x_{c}} \right) ^{2}$. On the other hand the DOS
can simply be estimated as $ N_{0} \approx x / W $, i.e. the dopant concentration divided by the bandwidth $W$,
which could only show the correct behavior stated above if the bandwidth were increasing faster than the dopant
concentration itself. It is quite unlikely that the DOS could have a concentration dependence that would make this
explanation consistent with the measurements, but one cannot entirely rule out that the resulting concentration
dependence is inconsistent with the Coulomb gap width as the DOS's concentration dependence is difficult to
ascertain and the arguments provided above simply provide a reasonable estimate.

\begin{figure}[tbh]
\centerline{\epsfig{figure=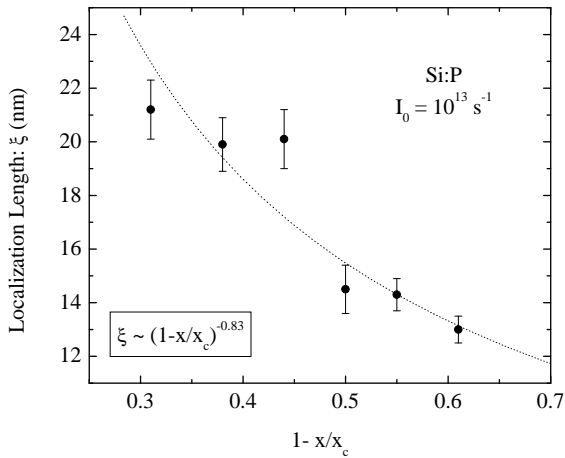,width=9cm}}\caption{Calculated
localization length plotted versus $1-\frac{x}{x_{c}}$ as
determined from the measured crossover frequency energies. Our
analysis entailed using an appropriate pre-factor for the overlap
integral \cite{Shklovskii}, $I_{0} = 10^{13} s^{-1}$ in the most
probable hop distance term, $r_{\omega} = \xi [\ell n ( 2 I_{0} /
\hbar \omega]$. \label{fig:SiP xi}}
\end{figure}

There is strong support however that the crossover energy is
determined by the Coulomb interaction energy though. By setting
the measured crossover energy scale equal to the Coulomb
interaction energy we are able to determine both the magnitude of
the localization length and the exponent with which it diverges as
a function of concentration. Using an appropriate pre-factor for
the overlap integral \cite{Shklovskii}, $I_{0} = 10^{13} \;
s^{-1}$, in the most probable hop distance term, $r_{\omega} = \xi
[\ell n ( 2 I_{0} / \hbar \omega]$, we find a localization length
dependence as shown in Figure \ref{fig:SiP xi}.

The localization length exponent as shown in the figure is close
to unity, the value originally predicted by McMillan in his
scaling theory of the MIT \cite{McMillan}, and the magnitude of
the localization length is reasonable. The choice of the prefactor
for the overlap integral may vary, but due to the fact that it is
located within a natural logarithm term, even an order of
magnitude variation greater or smaller than $10^{13}$ will result
in a change of approximately a factor of two in the localization
length. The consequent analysis for the localization length
exponent, namely after changing $I_{0}$ by an order of magnitude
both greater and smaller, still results in an exponent close to
unity for both cases. It must be mentioned that some experimental
results have indicated a localization length exponent closer to
$\frac{1}{2}$, but these were found on the metallic side of the
MIT using the zero temperature conductivity \cite{Hess}. These
results strongly point towards the Coulomb interaction energy as
being the energy scale at which the observed frequency dependent
crossover from ES to Mott-like hopping conduction occurs.

\begin{figure}[tbh]
\centerline{\epsfig{figure=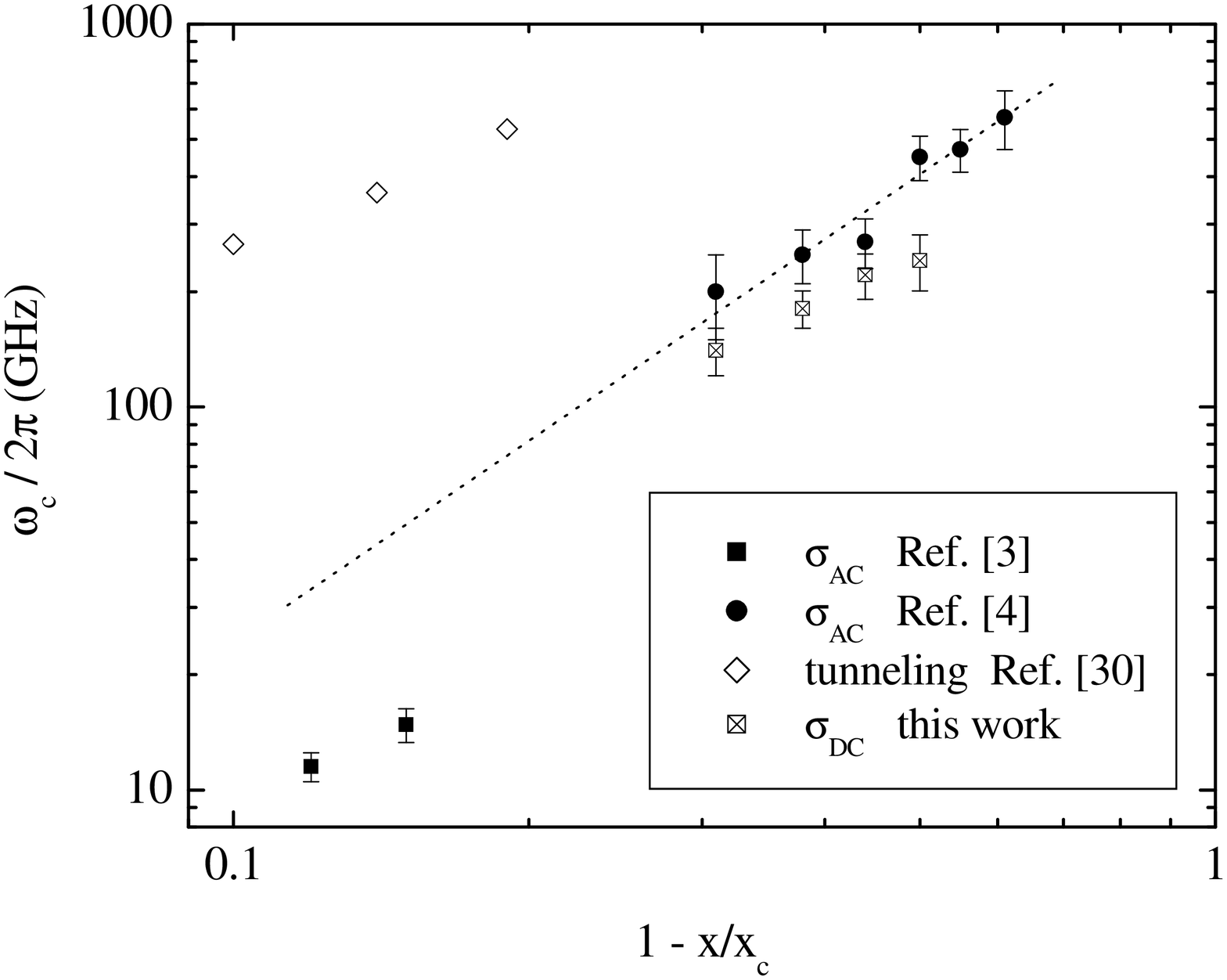,width=9cm}}\caption{Summary
of various experimentally determined crossover energy scales
determining the boundary between the interacting and
non-interacting states of an electron glass system. The data
represented by the open diamonds is adapted from tunnelling
measurements \cite{MLeePRB1999} and the data represented by the
full diamonds is adapted from AC conductivity network analyzer
measurements performed in a dilution refrigerator \cite{MLee01}.
Both measurements were done on Si:B, a disordered insulator
analogous to Si:P. Our measured crossover frequencies
\cite{Helgren2002} are represented by the full circles and our
experimentally determined crossover temperatures are converted to
frequency and are represented by the crossed
squares.\label{fig:All_crossover}}
\end{figure}

Figure \ref{fig:All_crossover} offers a summary of various
experimental data on different types of disordered insulating
systems, including our own, of measured DC and AC crossover
energies versus the normalized concentration parameter, $\left(1 -
\frac{x}{x_{c}} \right)$. The crossover energy for the various
experimental techniques are converted to frequency and displayed
in the figure as $GHz$. The data in the upper left of the figure
displayed as open diamonds is the Coulomb gap width in the single
particle DOS as determined by tunnelling measurements in Si:B
\cite{MLeePRB1999}, a disordered insulator analogous to Si:P.
Although the Coulomb gap, speculated by Lee et al. to set the
scale of the linear to quadratic crossover, was a smaller
re-normalized or "dressed" one, on this plot the single particle
gap sets an overall scale. One can see that the crossover energy
is well inside the single particle Coulomb gap for all energies.
The data in the lower left portion of the figure depicted by the
full diamonds are crossover frequencies as determined by Lee et
al. from AC conductivity measurements using a network analyzer and
a dilution refrigerator in order to achieve quantum limit-like
behavior on the same set of Si:B samples as the above tunnelling
data \cite{MLee01}. In their measurements, a sharp crossover
analogous to our own observations was seen. The solid circles in
Figure \ref{fig:All_crossover} represent the crossover frequencies
as determined from our frequency dependent conductivity
measurements on Si:P. Our data extends the observed range of the
sharp crossover deep into the insulating regime as the Si:B
measurements were closer to the critical concentration. The Si:P
and Si:B data do not fall on the same line. This may be due to
slight differences in these various disordered systems' dielectric
constants or localization lengths. Finally the open crossed
squares represent the crossover energies determined by us from our
DC conductivity measurements. As can be seen the DC crossover data
show a seemingly one-to-one correspondence to the observed
crossover energies with the AC data. This in itself is a curious
result, as the subsequent analysis of the DC data indicated that
the observed crossover was not consistent with a crossover from
Fermi glass Mott-like behavior to Coulomb glass ES-like behavior.

As a final note on the crossover data, recall that in the
literature (see Ref. 29) one finds experimental DC conductivity
data analyzed to indicate that in Si:P below a certain dopant
concentration, approximately $x/x_{c} = 80\%$, a "hard" Hubbard
gap opens up. Having observed appreciable AC conductivity at
frequencies below the purported hard gap energy, seems to rule out
the possibility of a "hard" Hubbard gap, as the optical
conductivity would be expected to drop precipitously to zero if
such a gap were present. In fact it seems that the picture set
forth by Efros and Shklovskii describing a crossover in the real
part of the complex conductivity from a linear to quadratic
dependence on frequency due to the effects of electron-electron
interactions, seems to hold from very close to the quantum
critical regime to deep within the disordered insulating electron
glass system at least qualitatively if not quantitatively. From
the expected form of $U(x)$, i.e. Eq. \ref{eq:U}, and the
reasonable behavior of the extracted $\xi$, it seems likely that
the crossover is governed by the Coulomb interaction strength of a
resonant pair as predicted by Efros and Shklovskii, but the
sharper than expected behavior still needs to be explained.

\subsection{AC Transport and Polarizability: The Coulomb Glass}

Having established the fact that indeed a crossover in the real
part of the frequency dependent complex conductivity occurs across
a broad range of dopant concentrations in an archetypical model
system of a disordered insulator, we wish to focus on the low
energy limit of this electron glass system, i.e. the Coulomb
glass. We will attempt to highlight successes and shortcomings of
the theoretical predictions by drawing on the results of both the
real and imaginary components of the complex conductivity of Si:P
and the previously reported results of another experimentally
determined Coulomb glass, namely NbSi \cite{Helgren2001}.
Comparing and contrasting the results of two Coulomb glasses, an
amorphous highly doped metal-semiconductor alloy, NbSi, and a
comparatively sparsely doped crystalline semiconductor, Si:P,
evidence for a general phase diagram of the Coulomb glass and the
quantum critical region is revealed relative to our experimentally
accessible window.

\begin{figure}[tbh]
\centerline{\epsfig{figure=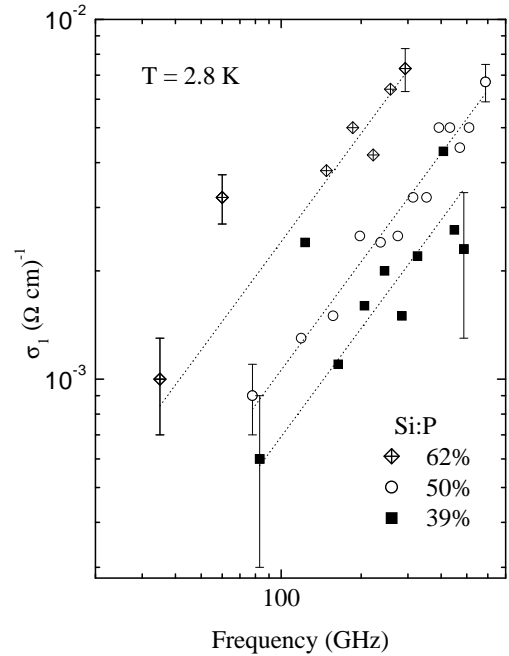,width=8cm}}\caption{The
low energy, linear in frequency portion of the real part of the
complex conductivity measured for three Si:P samples plotted
versus frequency. This data focuses on the Coulomb glass portion
of electrodynamic response. The dashed lines are forced linear
fits indicating self-consistency with the theory for a Coulomb
glass, namely a linear frequency dependence and an increase in the
magnitude of $\sigma_{1}$ with increasing dopant concentration.
\label{fig:SiP_Coulomb_Glass_s1}}
\end{figure}

\begin{figure}[tbh]
\centerline{\epsfig{figure=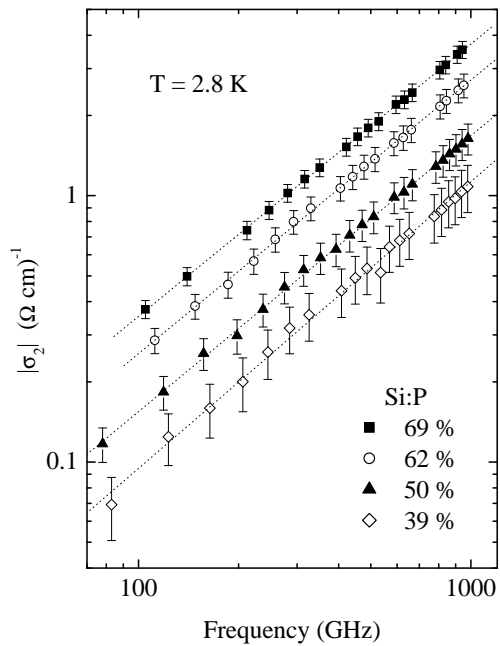,width=8cm}}\caption{The
low energy, linear in frequency magnitude of the imaginary part of
the complex conductivity measured for four Si:P samples plotted
versus frequency. This data focuses on the Coulomb glass portion
of electrodynamic response. The dashed lines are forced linear
fits indicating self-consistency with the theory for a Coulomb
glass, namely a linear frequency dependence and an increase in the
magnitude of $\sigma_{2}$ with increasing dopant concentration.
\label{fig:SiP_Coulomb_Glass_s2}}
\end{figure}

Figures \ref{fig:SiP_Coulomb_Glass_s1} and
\ref{fig:SiP_Coulomb_Glass_s2}  show the low energy magnitudes of
the real and imaginary components of the complex frequency
dependent conductivity plotted against frequency for a number of
Si:P samples. These figures focus on the low frequency Coulomb
glass data from the measurements of our experimentally accessible
region of the optical conductivity. The dashed lines are linear
fits to the data, highlighting two telltale characteristics of
Coulomb glass-like behavior, namely a linear dependence of both
components of the complex conductivity on frequency, and an
increase in the magnitudes of both $ \sigma_{1} $ and $ \sigma_{2}
$ with increasing dopant concentration. The error bars shown in
Figure \ref{fig:SiP_Coulomb_Glass_s1} are representative of all
the data point's error bars; they show a general trend of
increasing at lower frequencies both due to the logarithmic plot
and the increasing difficulty in measuring such low
conductivities. By contrast the error bars are much smaller and
there is little scatter of the data points from the linear fits in
Figure \ref{fig:SiP_Coulomb_Glass_s2} because due to the inherent
nature of an insulating system, the relative magnitude of the
imaginary component is greater than the real component of the
complex conductivity, and therefore falls in a range that can be
measured more accurately and precisely.

\begin{figure}[tbh]
\centerline{\epsfig{figure=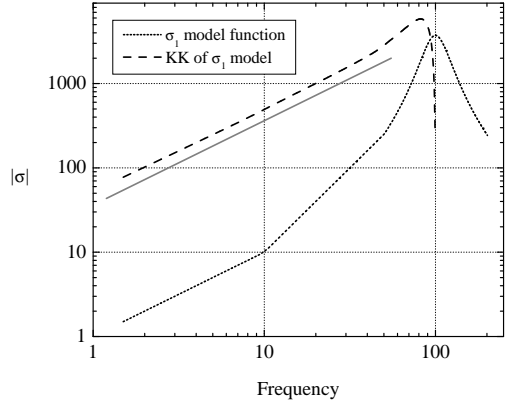,width=8cm}}\caption{A
model function of $\sigma_{1} (\omega)$, the dotted curve, and the
Kramers-Kronig calculated magnitude of $\sigma_{2} (\omega)$, the
dashed curve, on axes of arbitrary conductivity and frequency. The
model function for $\sigma_{1} (\omega)$ uses the Heaviside
function to change from a linear dependence on frequency to a
quadratic dependence at 10 on the arbitrary frequency scale. Above
100 on the arbitrary frequency scale a Lorentzian function is used
to simulate a peak in the optical conductivity spectrum. The
shaded gray line indicates a linear dependence. \label{fig:KK
analysis}}
\end{figure}

Closer inspection of Figure \ref{fig:SiP_Coulomb_Glass_s2} will
reveal that the $\sigma_{2}$ data is linear across the entire
measured spectral range, whereas the previous section clearly
established a crossover in $ \sigma_{1} $ as a function of
frequency. One might question whether this is physically
acceptable on the grounds of Kramers-Kronig (KK) compatibility.
This is established in Figure \ref{fig:KK analysis}. This figure
shows model data for $ \sigma_{1} $, plotted as the dotted line,
on an arbitrary conductivity axis that via proper utilization of
the Heaviside function mimics the Fermi glass to Coulomb
glass-like crossover, namely a change in the frequency dependence
of the real part of the complex conductivity from approximately
linear to approximately quadratic, at the value of 10 on the
arbitrary frequency scale. Above 100 on the arbitrary frequency
scale a Lorentzian function is used to simulate a peak in the
optical conductivity spectrum. The KK evaluated magnitude of
$\sigma_{2}$, resulting from the model data of $\sigma_{1}$, is
shown with the dashed curve. The imaginary component clearly
displays a linear dependence on frequency, as evidenced by the
shaded gray line, which is an aid to the eye for linear
dependence, even though there exists an abrupt kink in the real
part of the complex conductivity.

\begin{figure}[tbh]
\centerline{\epsfig{figure=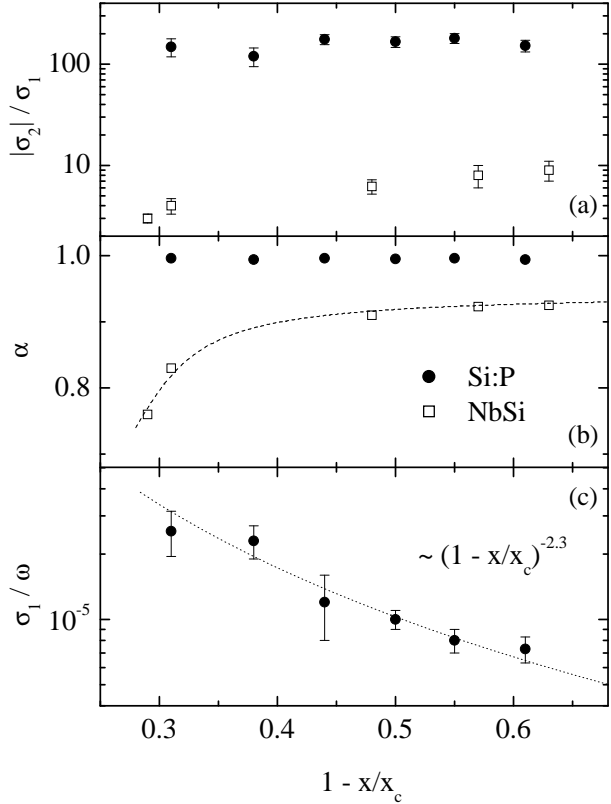,width=9cm}}\caption{Panel
(a) shows the ratio of the magnitude of $ \sigma_{2} $ to $
\sigma_{1} $ for the two Coulomb glasses measured in this work,
namely Si:P and amorphous NbSi. Panel (b) shows the calculated
powers of $\alpha$ as determined from Equation (\ref{eq:alpha}).
The dashed line through the NbSi data is a guide to the eye
showing an approach to the value $\frac{1}{2}$, as expected upon
approaching the quantum critical regime. Panel (c) shows the
divergence of the magnitude of $\sigma_{1}$ as a function of
concentration. \label{fig:SiP vs NbSi alpha}}
\end{figure}

Having established that below the crossover frequency, both components of the complex conductivity follow an
approximately linear frequency dependence allows us to analyze the full complex conductivity, $\hat{\sigma}$, using
the following Kramers-Kronig compatible form,
\begin{eqnarray}
    \hat{\sigma} = \sigma_{1} + i\sigma_{2} = A(i\omega)^{\alpha} \nonumber\\
    =A\omega^{\alpha} cos \left( \frac{\pi \alpha}{2}\right) + iA\omega^{\alpha} sin \left( \frac{\pi
    \alpha}{2}\right).
    \label{eq:KKsigma}
\end{eqnarray}
Such a dependence requires that the real and imaginary components follow the same power $\alpha$ over a broad range
of frequency and that this simple KK-compatible form only holds for $ \alpha \leq 1 $, thus this is perfectly valid
when analyzing the Coulomb glass regime. Given the functional form of the complex conductivity in Equation
(\ref{eq:KKsigma}), one can very accurately determine the power of $\alpha$ by taking the ratio of the magnitude of
$ \sigma_{2} $ and $ \sigma_{1} $ to obtain what is commonly known as the phase angle of the complex conductivity
\begin{eqnarray}
    \alpha = \frac{2}{\pi} \; tan^{-1} \left( \frac{|\sigma_{2}|}{\sigma_{1}} \right).
    \label{eq:alpha}
\end{eqnarray}
The ratio of $ |\sigma_{2}| / \sigma_{1}$ and the subsequent power of $\alpha$ was determined for each of the Si:P
samples and the results are summarized in Figure \ref{fig:SiP vs NbSi alpha}. A similar analysis was performed on
the insulating electron glass NbSi \cite{Helgren2001}, and the results from this Coulomb glass-like material are
shown in the figure for comparison purposes. Panel (a) in Figure \ref{fig:SiP vs NbSi alpha} shows the ratio of the
magnitude of the imaginary to the real component of the complex conductivity for both Si:P and NbSi.

It must be noted that the magnitude of the imaginary component plotted in the figure, for both Si:P and NbSi, is
that portion of the polarizability attributed to the interacting disordered electron system itself. Recall that the
experimentally determined quantity is the full dielectric constant, $\varepsilon_{1} = \varepsilon_{Si} +4 \pi
\chi$. The imaginary component of the complex conductivity plotted in the figure has been calculated from the
dielectric susceptibility portion $4 \pi \chi$, resulting from the electron-electron interactions via the
relationship $| \sigma_{2} | \propto (4 \pi \chi) \; \omega$. The ratio of $ |\sigma_{2}| / \sigma_{1}$ using an
imaginary component of the complex conductivity calculated from the full measured dielectric constant would
necessarily diverge as the dopant concentration decreased, i.e. deep in the insulating regime, because $ |
\sigma_{2} |$ would remain constant due to the static background dielectric constant from silicon itself, while $
\sigma_{1} $ would continuously decrease.

Returning to panel (a) in the figure, we first note that this ratio for Si:P remains large and approximately
constant across our range of dopant concentrations. From theory, one expects $|\sigma_{1}|$ to be approximately
equal to $|\sigma_{2}|$ to within a factor of 2-5 (with a reasonable estimate for $I_{0}$) as predicted by Efros
\cite{Efros1985}. Applied to Si:P, Eq. (\ref{eq:ESxover}) in the $\hbar\omega < U(r_{\omega})$ limit correctly
predicts a linear correspondence between $\sigma_{1}$ and $|\sigma_{2}|$, but the theory incorrectly predicts the
measured proportionality by at least a factor of thirty. The proportionality is closer for NbSi but has a
dependence on the doping concentration. This may be related to entering the quantum critical (QC) regime as
discussed below. The magnitude of the ratio becomes even larger and the discrepancy greater if the full dielectric
constant $\varepsilon_{1}$ is considered instead.

Panel (b) in Figure \ref{fig:SiP vs NbSi alpha} shows the power
$\alpha$ as determined by Eq. (\ref{eq:alpha}) for both the
amorphous NbSi and Si:P samples. The values for Si:P are
approximately equal to, but slightly less than one, consistent
with Figures \ref{fig:SiP_Coulomb_Glass_s1} and
\ref{fig:SiP_Coulomb_Glass_s2}. This indicates that the prefactor
of the real and imaginary components of the complex conductivity
for our samples of Si:P have the same concentration dependence
across our entire range of dopant concentrations.

Panel (c) in Figure \ref{fig:SiP vs NbSi alpha} shows the
magnitude of the real part of the complex conductivity for the
Si:P samples as the MIT is approached from the insulating side.
This demonstrates that the prefactor A as set forth in Equation
(\ref{eq:KKsigma}) can be written as a function of the normalized
concentration, i.e. $A \propto (1-\frac{x}{x_{c}})$ for Si:P. The
prefactor of the imaginary component can be analyzed in a similar
fashion and shows the same functional divergence as the real
component, i.e. $\sim (1-\frac{x}{x_{c}}) ^{-2.3} $. This
indicates that the real and imaginary components have the same
concentration dependence consistent with the simple Kramers-Kronig
form of the complex conductivity given in Equation
(\ref{eq:KKsigma}). The fact that across a broad dopant range on
the insulating side of the MIT we observe a similar divergence in
both components of the complex conductivity for Si:P is in
accordance with the concentration dependent theoretical
predictions for photon assisted hopping conductivity \cite{ES85},
for excitations both within and larger than the Coulomb gap energy
scale.

Still focussing on the Coulomb glass regime, the experimentally
determined concentration dependence of the prefactor A, namely $
(1-\frac{x}{x_{c}}) ^{-2.3} $, as shown in panel (c) of Figure
\ref{fig:SiP vs NbSi alpha}, can be used to determine which
theoretically predicted functional form for the real part of the
complex conductivity is more germane for our measured system.
Recall that for photon energies smaller than the Coulomb gap
width, i.e. excitations outside of the gap, the concentration
dependent terms are summarized by the following: $ \sigma_{1}
(\omega) = A \omega \propto \left(\frac{N_{0}^{2} \xi
^{4}}{\varepsilon_{1}} \right) \omega$, and that for the alternate
case, hopping conduction inside of the Coulomb gap one finds: $
\sigma_{1} (\omega) = A \omega \propto \varepsilon_{1} \omega$.
Having experimentally determined a parameterization for the
concentration dependencies for both the full dielectric constant,
$ \varepsilon_{1} \propto (1-\frac{x}{x_{c}}) ^{-0.4}$, and the
localization length, $\xi \propto (1-\frac{x}{x_{c}}) ^{-0.83}$,
we find that the functional form for the prefactor is better
described by photon-assisted hopping conduction occurring outside
of the Coulomb gap. We find that $ \xi ^{4} / \varepsilon_{1}
\propto (1-\frac{x}{x_{c}}) ^{-2.9}$, which is not in exact
agreement with our experimental findings for the prefactor (and in
fact the density of states, $N_{0}$ term would make this a
slightly steeper function), but this is in much better agreement
than the predicted concentration dependence of the prefactor for
photon assisted hopping conduction inside the Coulomb gap.

\begin{figure}[tbh]
\centerline{\epsfig{figure=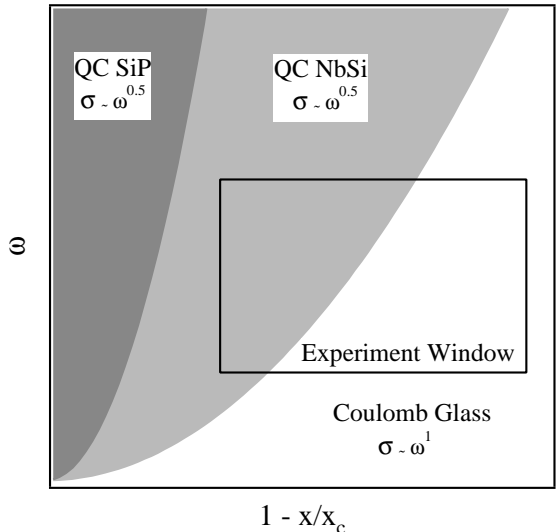,width=8cm}}\caption{A
schematic of a possible phase diagram indicating the relative
positions of the crossover energy scale to QC behavior vs.
concentration of Si:P and NbSi. The solid box indicates the
"window" from which our experimental results are derived.
\label{fig:SiP_NbSi_phase_diagram1}}
\end{figure}

\subsection{Quantum Critical Regime}

When approaching the MIT from the insulating side, the frequency
dependence is expected to cross over to the QC behavior
\cite{Henderson,Carini1998}, i.e. $\sigma_{1} \propto
\omega^{1/2}$ in NbSi, when the localization length $\xi$ becomes
comparable to the characteristic frequency dependent length scale,
the dephasing length, $\ell_{\omega}$ \cite{Sondhi}, as discussed
in Section II A. The crossover is not a phase transition and is
not expected to be sharp. Therefore looking at a fixed window of
frequencies, a broad, smooth crossover from $\omega \rightarrow
\omega^{1/2}$ would show an averaged power of the frequency
dependence similar to that measured for NbSi shown in the middle
panel of Figure \ref{fig:SiP vs NbSi alpha}. The fact that we see
an $\alpha \approx 1$ across our whole doping range in Si:P, but
an $\alpha$ that approaches $\frac{1}{2}$ in NbSi indicates that
the critical regime in Si:P is much narrower. Simple dimensional
arguments \cite{Kivelson} give a result similar to the
non-interacting case \cite{Shapiro1981} that the crossover should
be inversely proportional to the dopant DOS. The vastly smaller (a
factor of $10^{3}$) dopant density in Si:P relative to NbSi is
consistent with a narrower QC regime in Si:P as compared to NbSi.

A schematic outlining our experimental window and the boundaries
for the crossover from the insulating Coulomb glass regime to the
QC regime in both Si:P and NbSi versus a normalized concentration
is shown in Figure \ref{fig:SiP_NbSi_phase_diagram1}. The
differently shaded regions indicate the relative positions of the
crossover energy scale to QC behavior for Si:P and NbSi. As
mentioned above the crossover between the QC and the Coulomb glass
regimes occurs when the localization length scale becomes
comparable to the dephasing length scale, namely $ \xi \approx
\ell_{\omega} $, and from this relationship one finds that the
crossover frequency is inversely proportional to the DOS
\cite{Shapiro1981}. This crossover then is expected to be narrower
in Si:P as compared to NbSi due to its smaller DOS. The crossover
condition for the frequency in terms of the normalized
concentration is
\begin{eqnarray}
    \ \omega_{QC} \propto \ell_{0}^{z}\left(1-\frac{x}{x_{c}}\right)^{z\nu}
    \label{eq:RegimeBoundary}
\end{eqnarray}
where the subscript QC refers to the quantum critical regime. The
functional dependence is expected to be universal but the
prefactor can vary from system to system. This then defines a
condition for the crossover from QC to insulating glass-like
behavior that depends only on the dynamical scaling exponent, $z$,
and the localization length exponent, $\nu$, that is consistent
with experimental evidence. For the case of $ z \approx 2 $, and $
\nu \approx 1 $ as evidenced in NbSi \cite{Henderson} and Si:P
\cite{Helgren2002} respectively, the boundary would then have an
approximately quadratic dependence on the normalized concentration
as shown in Figure \ref{fig:SiP_NbSi_phase_diagram1}.

\section{Discussion}

In analogy to the three phases of matter we are accustomed to dealing with in elementary physics, the solid, liquid
and gas phases, a collection or assembly of electrons can be classified in a similar manner depending on the
electrodynamic response of the system, each possessing an accompanying theoretical model. The room temperature
response of a metal upon applying an electric field is well described by the Drude model. Here the electron system
can be thought of as an electron gas with a temperature dependent mean free path and scattering time. An electron
liquid would naturally be a system of electrons obeying Landau's theory of Fermi liquids
\cite{Landau1957,Landau1959}. This successful theory has distinct hallmarks in the low energy limit. We shall
investigate this low energy behavior more closely in the following section by comparing it to the low energy
behavior of the materials we studied, generally classified as electron glasses.

The poorly understood and infrequently studied electron glasses
might be considered a genre of the final phase mentioned above,
namely an electron solid, where the electronic states are
localized due to disorder. The amorphous nature of the glassy
state connotes disorder and hence this appellation was chosen to
describe this type of assembly of electrons. Via our unique and
versatile AC conductivity measurement techniques, and by
investigating a series of disordered insulating systems proximal
to a MIT, we have sufficient experimental evidence for electron
glass systems to propose a classification scheme, i.e. a
'taxonomy' for the electrodynamic response of electron glasses,
along with a phase diagram containing appropriate crossover
boundaries for the electron glass, Coulomb glass and quantum
critical regimes.

\subsection{On the Nature of the Coulomb Glass}

One of the fundamental reasons for undertaking our investigations
of these electron glasses is the surprising fact that even in a
material as thoroughly studied as doped bulk silicon (due to the
relevance of such research for the semiconductor industry), there
exists no clear consensus as to the ground state and the nature of
the low energy excitations in these disordered insulating systems.
Using the frequency dependent complex conductivity we wished to
probe the question as to whether or not the Coulomb interactions
fundamentally change the ground state of the system. A philosophy
that persevered for many years was that in analogy to the Fermi
liquid theory, the theory describing the excitations at
arbitrarily low energies in a system localized due to disorder,
would indeed not be qualitatively influenced by electron-electron
interactions \cite{Anderson1970}. The theory created under this
pretense was the one set forth by Mott \cite{Mott}, which gives an
approximately quadratic frequency dependence of the real component
of the complex conductivity. The coining of the appellation Fermi
glass to describe this non-interacting disordered electron system
was a direct consequence of the belief that there existed an
analogy between the low energy electrodynamics of a Fermi liquid
and this Fermi glass. Later, theoretical work and subsequent
investigations of the DC conductivity lent credence to the import
of long range interactions stemming from negligible screening of
the low energy excitations, e.g. the work by Pollak
\cite{Pollak1970,Pollakbook85} and by Efros and Shklovskii
\cite{ES85}.

In 'many' systems in solid state physics, this claim certainly is
true, namely that the effects of correlations become simpler at
low frequency or energy. The canonical example, as stated, is the
Fermi liquid. The heavy-electron compounds such as $CeAl_{3}$ and
$UPt_{3}$ are materials that have strongly correlated, yet
delocalized charge carriers, and in the low energy limit, they
follow the non-interacting Fermi liquid theory with renormalized
parameters such as the effective mass and the relaxation rate
\cite{Awasthi}. For these heavy-fermion systems there is a one to
one correspondence between the low-lying excitations and those in
a nearly free electron gas when the parameters of the latter,
namely the mass and scattering rate, are properly renormalized to
take into account the correlations. The name 'heavy-electron'
stems from this effective enhanced mass (often several hundred
times) which is a direct consequence of the electron-electron
interactions, but nevertheless the theory describing the
electrodynamic response of these systems is a non-interacting one.

On the other hand, our experimental data for both NbSi and Si:P,
two variants of electron glass systems, show Coulomb glass-like
behavior, i.e. a theory dependent on interactions, in the low
energy frequency dependent electrodynamic response. Experiments
done by another group on an electron glass system \cite{MLee01},
namely Si:B, also show a linear in frequency power law dependence
of the $T=0, \; \omega \rightarrow 0$ conductivity supporting and
verifying the import of electron-electron interactions in the
low-energy response of these disordered systems. While this work
on Si:B reported on only two concentrations of dopant very close
to the MIT, our further investigation across a broad range of
dopant concentrations in Si:P can confirm that this linear power
law dependence continues deep into the insulating regime. Thus we
have clear evidence that an interacting theory, namely that of a
Coulomb glass, dependent on electron-electron correlations, is
correct at the lowest experimentally accessible energy scales in
these electron glass systems, in opposition to the original
thoughts on the matter. Thus the Coulomb glass is a fundamental
example in solid state physics as the non-interacting functional
form is not recovered in the asymptotically low energy limit.

\subsection{The Taxonomy of an Electron Glass}

Now we focus on the electron glass phase as a whole with the
intent of setting forth a universal framework for the AC response
of such disordered insulating systems. In summary, we have
observed Coulomb glass-like behavior across our entire range of
doping concentrations in Si:P and via a comparison with NbSi have
arrived at a boundary condition consistent with our experimental
data for the crossover from QC to Coulomb glass-like behavior,
namely Equation (\ref{eq:RegimeBoundary}). We have also
established that a crossover from Fermi glass-like to Coulomb
glass-like behavior exists, as is manifest in the experimental
data of the real part of the complex frequency dependent
conductivity, not only close to the MIT \cite{MLee01}, but also
deep into the insulating electron glass regime. Employing these
phenomenological results, namely that the power laws seen in the
frequency dependent complex conductivity are qualitatively
consistent with the three expected types of frequency dependent
behavior theoretically postulated for the electron glass regime,
we propose that the real component of the frequency dependent
complex conductivity of these disordered insulators can be
parameterized in general by
\begin{eqnarray}
    \ \sigma_{1} \propto \omega ^{\alpha ( \omega, x )}
    \label{eq:Parameterized Sigma1}
\end{eqnarray}
where the power $\alpha$ depends on the frequency and the dopant
concentration. This classification scheme, or taxonomy, based on $
\alpha ( \omega, x) $, captures the essential and germane physics
of the electrodynamic response of electron glasses.

\begin{figure}[tbh]
\centerline{\epsfig{figure=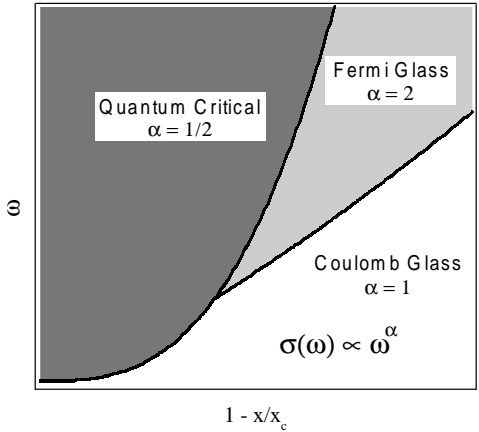,width=9cm}}\caption{This
schematic of a possible phase diagram for an electron glass shows
the regime boundaries for the QC, the Coulomb glass and the Fermi
glass regimes in a two dimensional space of frequency versus
normalized concentration $\left( 1 - \frac{x}{x_{c}} \right)$.
\label{fig:Electron_Glass_phase_diagram}}
\end{figure}

Figure \ref{fig:Electron_Glass_phase_diagram} summarizes our
proposed phase diagram and taxonomy on a plot of frequency versus
the normalized concentration $(1-\frac{x}{x_{c}})$. Recall that
the criterion for the regime boundary condition for the crossover
from Coulomb glass or Fermi glass-like behavior to the region
where QC dynamics dominated was determined by Eq.
(\ref{eq:RegimeBoundary}). For a $z=2$ theory and a localization
length exponent $\nu = 1$, this crossover would go as $\omega
_{qc} \propto (1-\frac{x}{x_{c}})^{2}$ with a prefactor dependent
on the dopant DOS. For the crossover from Fermi glass to Coulomb
glass-like behavior, shown on the right hand side of the figure as
the light gray and white regions respectively, we found that the
crossover energy was more likely to be determined by the Coulomb
interaction energy as opposed to the Coulomb gap width using
concentration dependent arguments. Either way our experimentally
determined crossover had the following concentration dependence,
$\omega _{c} \propto (1-\frac{x}{x_{c}})^{1.65}$. Here we must
note that though the curve depicting this latter crossover, as
shown in the figure, might intersect the QC crossover curve
(depending on the material's energy scales), the QC regime will
always take precedence. Recall that this crossover energy stems
from a length scale argument that quantum critical behavior occurs
when the electronic localization length, $\xi$, is greater than
the dephasing length, $\ell _{\omega}$. Ergo since both the Fermi
glass and Coulomb glass behavior only occur in the opposite case,
e.g. $\xi < \ell _{\omega}$, the length scale dependent crossover
condition for QC behavior always dictates the low energy, close to
critical behavior.

The classification of the power of $ \sigma_{1} \propto \omega ^{ \alpha( \omega, x)} $, e.g. $\alpha = 1/2$ for QC
dynamics, $\alpha = 1$ for Coulomb glass behavior and $\alpha = 2$ for Fermi glass behavior, provides more than
just a taxonomy for these disordered insulating systems. For the latter two regimes, with powers 1 and 2
respectively, these numbers also represent the dimensions of phase space changing directly as $\hbar \omega$ that
these systems have access to for the creation of a real excitation.

\section{Conclusion}

In summary, we have presented results of our frequency dependent
complex conductivity investigations into correlated electron
phenomena in the electrodynamics of the electron glass Si:P. The
insulating side of the MIT displayed the expected power law
dependencies for a Fermi glass and a Coulomb glass system with the
latter being the low energy limit in our experimentally accessible
window. In comparison with another electron glass, NbSi, with a
distinctly different DOS, wherein we see a crossover from Coulomb
glass-like behavior to quantum critical behavior, we have
formulated a phase diagram based on phenomenological observations
of all the predicted states for the insulating side of the MIT.
The taxonomy for the electron glass that we have presented, is an
attempt to both capture the essence of the physics and allow for a
general description of the low energy frequency dependent
electrodynamics of electron glasses.

A number of issues have also been left unresolved by this work.
First of all, the apparent sharpness of the crossover in the real
part of the frequency dependent conductivity cannot be explained
by the current model of Efros and Shklovskii \cite{ES85}. Second,
there certainly also exist further misgivings with respect to the
theories concerning the DC conductivity. For instance, the DC
conductivity data does not seem to fit the Mott or ES predictions
particularly well  for our doping levels well into the insulating
phase as indicated by the unphysical characteristic temperatures
derived therefrom. Also, we have experimentally observed a
crossover energy in the DC conductivity that is quite comparable
to the crossover energy seen in the frequency dependent
conductivity, and there is no reason within the canonical theory
that there should be a correspondence in the AC and DC energy
scales. Thirdly the experimentally determined ratio of the
imaginary to the real component of the complex conductivity shows
a deviation from the predicted ratio. Should a subsequent
re-analysis of the theories pertaining to the electrodynamics of
these electron glasses be undertaken, one should certainly include
a self-consistent dynamic dielectric constant, and in a correct
description, the proper ratios and relative sharpness of the
crossovers must be manifest.

We are also hopeful that this work will plant the seed for further
future investigations of other disordered insulating systems. For
instance the samples used in these investigations were nominally
uncompensated. Purposeful compensation will naturally redistribute
many relevant energy scales. It would be of fundamental importance
to investigate the veracity of the proposed taxonomy of the
electron glass with respect to compensated systems. One must also
note that our phase diagram stemmed from phenomenological
observations of the three types of electron glass behavior, namely
Fermi glass, Coulomb glass and quantum critical, but we would be
remiss if we neglected to emphasize that we did not physically
observe a crossover from Fermi glass to quantum critical behavior,
just the other crossovers that are possible in the taxonomy phase
diagram. Needless to say, such an observation would be of great
import in fully understanding electron glass behavior. Finally we
have seen that in our experimental window, the original
expectations set forth in the literature \cite{Anderson1970} have
proven to be not realized, in that at our lowest accessible
frequencies, an interacting theory is the germane one. This does
not however rule out a subsequent change in the conduction
mechanism at even lower frequencies, such as non-linear effects.

\section{Acknowledgements}

We would also like to thank S. Kivelson, B. Shklovskii and E.
Abrahams for useful discussions and also N. F. Mott who introduced
one of us (GG) to the subject. This research was supported by the
National Science Foundation grant DMR-0102405.

\end{document}